\documentclass[%
 reprint,
preprintnumbers,
 amsmath,amssymb,
 aps,
]{revtex4-2}

\usepackage{graphicx}
\usepackage{dcolumn}
\usepackage{bm}
\usepackage{appendix}
\usepackage{hyperref}

\begin{document}

\preprint{CERN-TH-2025-218}

\title{QCD String Axions and $M$-theory}

\author{Bobby S. Acharya}
\email{bacharya@ictp.it}
\affiliation{Abdus Salam International Centre for Theoretical Physics, Strada Costiera 11, 34151, Trieste, Italy}
\author{Ethan Torres}
 \email{ethan.martin.torres@cern.ch}
\affiliation{Theoretical Physics Department, CERN, 1211 Geneva 23, Switzerland}%




\date{\today}

\begin{abstract}
In $SU(N)$ Yang-Mills theory without matter, there exist stable long electric fluxtube strings which carry a 1-form symmetry charge. Over the past decade or so, there has been increasing evidence from lattice calculations that the worldsheet theories of such QCD strings contain a massive pseudoscalar (axion), at least when $N\geq 3$. This has so far been puzzling from the perspective of holographic realizations of strings in confining gauge theories. In this note, we will show how such axions appear naturally in the realization of 4D $\mathcal{N}=1$ Super-Yang-Mills (SYM) from $M$-theory spacetimes in which the extra dimensions are modeled by certain complete metrics of $G_2$-holonomy. This picture predicts that QCD string axions exist only if the gauge group is $SU(N\geq 3)$, $SO(4N+2)$, or $E_6$ (or a quotient/double-cover thereof), and is absent from the spectrum of stable QCD strings if the gauge group is $SU(2)$, $SO(2N+1)$, $SO(4N)$, $Sp(N)$, or $E_7$. We argue why we expect this pattern to persist for non-supersymmetric Yang-Mills strings, at least for large $N$, something which could be tested in future lattice studies of QCD strings for gauge algebras of B, C, D, and E-type.
\end{abstract}

\maketitle


\section{Introduction}
Since the dawn of string theory, it has been proposed that the strong interactions can be described in terms of strings. Linearly rising Regge trajectories and $s$- and $t$- channel duality in 2 $\rightarrow$ 2 scattering are indeed approximate features of real world hadron physics and were key motivations behind the bevy of activity in the early pre-QCD string theory literature (for a historical account see \cite{Green:1987sp}). Although QCD has taken the place as the fundamental theory of the strong interactions, many features of the long-distances physics is well-described by the existence of long chromo-electric fluxtube strings (QCD strings) in the spectrum. This includes the linear behavior of the quark anti-quark potential, as well as the phenomenological successes of the Lund string model. Given a gauge group $G$ with Lie algebra $\mathfrak{g}$ \footnote{We will assume that $\mathfrak{g}$ is simple throughout.}, the creation/annihilation operators for strings are Wilson lines in representations of $\mathfrak{g}$, and the fact that such Wilson lines obey an area-law at long-distances means that the QCD strings they create includes a Nambu-Goto term in their 2D effective action.

Understanding finer details of these effective actions is an active area of research and one of the most exciting developments in the past decade or so has been the inclusion of a massive pseudoscalar particle, often called a QCD string axion, in the worldsheet effective theory \cite{Athenodorou:2010cs,Dubovsky:2013gi}. This particle is characterized by the fact that it is odd under transverse spatial reflections. The necessity of such a particle was motivated to match lattice calculations and evidence for its existence (at least for the cases of $G=SU(N\geq 3)$) have only grown \cite{Athenodorou:2017cmw, Gaikwad:2023hof, Athenodorou:2024loq, Sharifian:2025fyl}. A (potential) intuitive explanation, due to Dubovsky \cite{Dubovsky:2018vde} (see also \cite{Gabai:2025hwf}), is that if one thinks of the massless 2D scalars parametrizing the normal directions of a long QCD string, $X^{i=1,2}$, as coming from the UV variable $\mathrm{Tr}\exp(i\oint A) (F_{3i})$, then the pseudoscalar arises from a UV variable $\mathrm{Tr}\exp(i\oint A) (F_{12})$. Under the transverse rotational group, $O(2)_\perp$, these operators are in the representations $J^{\mathsf{R}_1}=1^+$ and $0^-$, respectively. So far, the presence of the pseudoscalar has posed a puzzle from the point of view of string theory/holographic realizations of strings in confined gauge theories where the spin-0 spectrum appears to naively only include scalars \textit{even} under $\mathsf{R}_1$.

Within the framework of string/$M$-theory there exist a variety of four dimensional models containing confining strings (for a sampling of classic references see \cite{Witten:1997ep, Hanany:1997hr, Maldacena:2000yy, Klebanov:2000hb, Vafa:2000wi, Acharya:2000gb, Atiyah:2000zz, Acharya:2001hq}). In general, these models tend to have regions of parameter space which are not present in the Yang-Mills theory and can formally be continued beyond infinite coupling: the confining strings exist explicitly in these models and can be described in regimes where the bulk spacetime is smooth and low energy field theory is a good approximation.
In this note, we will investigate confining strings in particular $M$-theory realizations of the minimal ($\mathcal{N}=1$) supersymmetric version of Yang-Mills theory which includes a massless adjoint Majorana fermion. They are described by $M$-theory spacetimes in which the seven extra dimensions of space are modeled by certain spaces admitting metrics with $G_2$-holonomy and were first studied in this context in   \cite{Acharya:2000gb, Atiyah:2000zz, Aganagic:2001ug, Aganagic:2001jm}. The $G_2$-holonomy ensures spacetime supersymmetry and one has smooth solutions in which there is an unbroken 1-form symmetry, confining strings charged under the symmetry and a mass gap; the latter feature arises by showing that there are no $L^2$-normalizable massless modes.

In more detail, these models arise from the Bryant-Salamon metrics \cite{10.1215/S0012-7094-89-05839-0} on the total space of the
spinor bundle over a round 3-sphere, $\Sigma S^3$. These are geodesically complete, asymptotically conical solutions to the vacuum Einstein equations. $\Sigma S^3$ is topologically $\mathbb{R}^4 \times S^3$ since the spinor bundle is trivial. Finally, the actual models arise by taking discrete quotients by finite $ADE$ subgroups, $\Gamma_{ADE}$, of various $SU(2)$-isometry groups of the spacetime and different models arise depending on whether or not $\Gamma_{ADE}$ acts on the $\mathbb{R}^4$ factor or the $S^3$ factor. In the former case, the quotient space has orbifold singularities, whilst in the latter it is smooth, but it turns out \cite{Atiyah:2001qf} that there is a smooth parameter space which allows one to interpolate between these and therefore from perturbative $\mathcal{N}=1$ Super-Yang-Mills (SYM) to these other models.
4D $\mathcal{N}=1$ SYM is known to confine at low-energies and much its physics is neatly captured in $M$-theory language because of the flow from the singular $G_2$-geometry to a smooth $G_2$ manifold $\Sigma(S^3/\Gamma_{\mathfrak{g}})$ \cite{Atiyah:2000zz, Acharya:2000gb, Aganagic:2001ug, Aganagic:2001jm}. Domain walls separating vacua with different gaugino condensate phases, are neatly captured by M5 branes wrapping the base 3-cycle of $\Sigma(S^3/\Gamma_{\mathfrak{g}})$. This picture was in fact crucial in proposing the low-energy description of such $\mathcal{N}=1$ SYM domain walls \cite{Acharya:2001dz}. QCD strings, on the other hand, arise from M2-branes wrapping non-contractible circles of $\Sigma(S^3/\Gamma_{\mathfrak{g}})$ \cite{Acharya:2001hq} and have received relatively little study.

Our main task will be to study the worldsheet theory of these strings, viewed as a circle compactification of the M2-brane worldvolume wrapping the cycles of $\Sigma(S^3/\Gamma_{\mathfrak{g}})$.
While a full description of the dynamics is outside of the scope of this paper,
we do identify a pseudoscalar mode in the M2-brane sigma model by understanding its transformation properties under spatial reflection. A crucial fact we use is that spatial reflection of a 4D spatial coordinate in the $M$-theory language, $\mathsf{R}^{11D}_{i=1,2,3}$, is what a 4D gauge theorist would call $\mathsf{C}\mathsf{R}^{4D}_{i=1,2,3}$ where $\mathsf{C}$ the usual charge conjugation \cite{Cvetic:2021maf}. We will show that this only occurs when the gauge algebra is
\begin{equation}\label{eq:summaryintro}
  \mathfrak{g}=\mathfrak{su}(n\geq 3), \; \mathfrak{so}(4k+2), \; \mathrm{or}\; \mathfrak{e}_6
\end{equation}
In such cases, the field theorists notion of a spatial reflection must include a discrete isometry of the $G_2$-manifold, $\mathsf{R}_{int}$, in order to undo the implicit action of $\mathsf{C}$, i.e.
\begin{equation}\label{eq:4d11drefl}
  \mathsf{R}^{4D}_{i}=\mathsf{R}^{11D}_{i}\circ \mathsf{C}=\mathsf{R}^{11D}_{i}\circ \mathsf{R}_{int}.
\end{equation}
This makes at least one of the M2-brane sigma model directions odd under $\mathsf{R}^{4D}_{i}$ which can thus be identified as a pseudoscalar.
\footnote{The QCD string axion is also expected to be odd under
$\mathsf{C}\mathsf{R}^{4D}_3(=\mathsf{R}_{3}^{11D})$ (recall that we take the long string spatial direction to be $x_3$) but if it is periodic, as
will be the case from our $M$-theory prediction, this quantum number is less meaningful since one
can exchange $\mathsf{C}\mathsf{R}^{4D}_3$-even/odd excitations upon dualizing. For instance, one can dualize a periodic $a$ to a periodic
$\tilde{a}$ via $da=*d\tilde{a}$.
}.
Notably, this precludes the existence of QCD string axions for the cases \footnote{We did not include $\mathfrak{e}_8$ as this theory has no stable QCD strings in the first place due to the fact that $Z(E_8)=1$.}
\begin{equation}
    \mathfrak{g}=\mathfrak{su}(2), \; \mathfrak{sp}(N), \; \mathfrak{so}(2N+1), \; \mathfrak{so}(4N), \; \mathfrak{e}_7.
\end{equation}

This note is organized as follows. We first review the relevant $G_2$-holonomy spacetimes in Section \ref{sec:review}, and the physics of $M$-theory on such manifolds in Section \ref{sec:g2flopconfinement}. Section \ref{sec:versus} presents our main argument for why \eqref{eq:4d11drefl} holds only for the gauge algebras listed in \eqref{eq:summaryintro}, Section \ref{sec:ccisometries} classifies the possible choices of $\mathsf{R}_{int}$, and Section \ref{sec:QCDstringworldsheet} analyzes the implications for the QCD string worldsheet theories. In Section \ref{sec:nonsusy}, we comment on the implications of our results for strings in non-supersymmetric Yang-Mills theory. We then state our conclusions where we mention that mismatches of the form \eqref{eq:4d11drefl} are expected to arise in a number of string theory/holographic realizations of confining gauge theories. Appendix \ref{app:isom} includes technical details on isometries of the $G_2$-holonomy spaces of interest.

\section{Review: $\mathcal{N}=1$ Yang-Mills from $G_2$-manifolds}\label{sec:review}
Four dimensional $\mathcal{N}=1$ supersymmetric Yang-Mills theory arises from $M$-theory spacetimes of the form
$M^{10,1} = (\Sigma S^3)/\Gamma  \times \mathbb{R}^{3,1}$ with a spacetime metric which takes the form $g(M^{10,1}) = g(\Sigma S^3) + \eta $, with $g(\Sigma S^3)$ the Bryant-Salamon metric on $\Sigma S^3$ and $\eta$ the Minkowski metric.
The Bryant-Salamon metrics are complete, $G_2$-holonomy metrics which are asymptotic to cones with cross-section $S^3 \times S^3 \subset \Sigma S^3 = \mathbb{R}^4\times S^3$. There are three such metrics corresponding to three distinct ways to resolve the conical singularity using three homologically distinct $S^3$'s $\subset S^3 \times S^3$. We label these with a subscript $a,b$ or $c$. The Bryant-Salamon
metrics, which have one real parameter $r_a$, on $\Sigma S^3_a$ can be written as
\begin{widetext}
\begin{equation}\label{eq:sigmas3metric}
  ds^2=\frac{dr^2}{1-(r_a/r)^3}+\frac{r^2}{24}da^2+\frac{r^2}{72}\left(1-\left(\frac{r_a}{r}\right)^3\right)(2db^2+2dc^2-da^2)
\end{equation}
\end{widetext}
where $a,b,c\in SU(2)_{a,b,c}\equiv S^3_{a,b,c}$ such that $abc=1$, and the radial coordinate $r$ has range $[r_a,\infty]$ \cite{10.1215/S0012-7094-89-05839-0}. We have used the notation of \cite{Atiyah:2001qf} where $da^2\equiv -\mathrm{Tr}(a^{-1}da)$, and similarly for $db^2$ and $dc^2$. Note that $r_a$ controls the radius of the minimal volume $S^3_a$ which is the zero section of the spin bundle, which is manifest in \eqref{eq:sigmas3metric}. When $r_a>0$, this metric has an $(SU(2)^3/\mathbb{Z}_2) \rtimes \mathbb{Z}_2$ isometry group, which in the above coordinates acts as \cite{Atiyah:2001qf}:
\begin{align}\label{eq:isometryactions}
  (u,v,w)\in SU(2)^3: & \; \; a\rightarrow uav^{-1}, \; b\rightarrow vbw^{-1}, \; c\rightarrow wcu^{-1} \\
  \iota_2\in \mathbb{Z}_2\subset S_3: \; \; & (a,b,c)\rightarrow (a^{-1},c^{-1},b^{-1}).
\end{align}
There is a $\mathbb{Z}_2$ quotient of the $SU(2)^3$ factor due to the fact that the element $u=v=w=\mathrm{diag}(-1,-1)$ acts trivially on $a$, $b$, and $c$. When $r_a=0$, this isometry group is enhanced to $(SU(2)^3)/\mathbb{Z}_2 \rtimes S_3$ which includes the additional generator $\iota_3\in \mathbb{Z}_3\subset S_3: \; \; (a,b,c)\rightarrow (b,c,a)$. Note that these isometries all preserve the orientation of $\Sigma S^3_{a,b,c}$ \footnote{There exist orientation reversing isometries of $S^3_a\times S^3_b\times S^3_c$ such as $a\rightarrow a^{-1}$, and $a \leftrightarrow b$, but none of these are preserved after imposing $abc=1$.}.

At the $r_a=0$ point in the geometric parameter space, the $S^3_a$ at the center contracts to a point and we have a conical singularity. At this point
one can switch on a parameter $r_b>0$ to a non-zero minimal volume to $S^3_b$ meaning that the $G_2$-manifold is now $\Sigma S^3_b$. The metric is then given by the permutation of \eqref{eq:sigmas3metric} by $\iota_3$ and the replacement $r_a\rightarrow r_b$. This process in the geometric parameter space is known as a \textit{$G_2$-flop transition} \cite{Atiyah:2000zz}.

By considering specific discrete quotients of the above three $G_2$-holonomy manifolds gives rise to $M$-theory spacetimes which geometrically realize 4D $\mathcal{N}=1$ SYM with (simply-laced) gauge algebra $\mathfrak{g}$. These arise from quotients by by $\Gamma_{\mathfrak{g}}\subset SU(2)_u$ where the subscript of $\Gamma_{\mathfrak{g}}$ denotes the ADE label in the classification of discrete subgroups of $SU(2)$. The gauge \textit{group} $G$ is ultimately related to choosing boundary conditions at $r=\infty$ \cite{Albertini:2020mdx, Khlaif:2025jnx} and because these various choices will not play a role in this note, we will simply assume $G$ to be \textit{simply-connected} throughout \footnote{One change between these cases relevant to QCD strings is that if we take $G$ to be simply connected, then the strings will have 1-form global symmetry charge $Z(G)^{(1)}$, while if we instead choose $G/Z(G)$, then the QCD strings would be charged under a $Z(G)^{(1)}$ \textit{gauge} symmetry.}. Giving a concrete presentation of the $A_{N-1}$ cases, we obtain $\mathfrak{su}(N)$ SYM theory by quotienting by a $\mathbb{Z}_N\subset SU(2)_u$ action generated by
\begin{equation}\label{eq:udef}
  u_0=\begin{pmatrix}
      \zeta_N & 0 \\
      0 & \zeta^{-1}_N
    \end{pmatrix}
\end{equation}
The $SU(2)$ element $a$ can be written in terms of complex coordinates where $|z_{1a}|^2+|z_{2a}|^2=\frac{1}{36}(r^2+(\frac{r^3_a}{2r}))$ and
\begin{equation}\label{eq:acoord}
a=\frac{6}{\left(r^2+(\frac{r^3_a}{2r})\right)^{1/2}}\begin{pmatrix}
      z_{1a} & -z^*_{2a} \\
       z_{2a} & z^*_{1a}
    \end{pmatrix}.
\end{equation}
which is normalized so that $\mathrm{det}(a)=1$. As will be convenient later, one can also write $a$ as a unit quaternion:
\begin{equation}\label{eq:unitquart}
  a=(z_{1a}+jz_{2a})/||z_{1a}+jz_{2a}||.
\end{equation}

Quotienting by $u_0$ then identifies
\begin{equation}\label{eq:znaction}
  (z_{1a},z_{2a})\sim (\zeta_N z_{1a}, \zeta^{-1}_Nz_{2a})
\end{equation}
which defines the Bryant-Salamon orbifold $\Sigma (S^3_a/\mathbb{Z}_N)$ which will be relevant to the IR physics. Relevant to the UV physics will be the same $\mathbb{Z}_N$ quotient as in \eqref{eq:znaction} but on the flopped geometry. We denote this geometry by $\Sigma_{\mathfrak{su}(N)}S^3_b$ where now the base of the flopped geometry, $S^3_b$, is invariant under \eqref{eq:znaction} and leads to a codimension-4 fixed locus where the 7D $\mathfrak{su}(N)$ gauge degrees of freedom live. The more general quotients $\Gamma_{\mathfrak{g}}\subset SU(2)_u$ are denoted $\Sigma_{\mathfrak{g}}S^3_b$.

\section{Review: $G_2$-Flop and Confinement}\label{sec:g2flopconfinement}



A key tool in understanding the physics of $M$-theory on such $G_2$-orbifolds is the aforementioned $G_2$-flop transition. Consider $M$-theory on $\Sigma_{\mathfrak{g}} S^3_b$ with fixed $V_b=:M^{-3}_{KK}$ which we initially take to be much larger than the Planck length scale $M^{-1}_{P}$. At low energies, the 11D supergravity sector decouples and the physics is described by 7D SYM with gauge group $G$ reduced on $S^3_b$ which yields a 4D gauge coupling
\begin{equation}
    g^2_0:=g^2(M_{KK})=\frac{M^3_{KK}}{M^3_P}
\end{equation}
Here $g^2(\mu)$ denotes the 4D gauge coupling measured at an energy scale $\mu$, and below this KK-scale, there are no chiral multiplets which means that the 1-loop running (exact in the holomorphic scheme) is
\begin{equation}
    \frac{1}{g^2(\mu)}=\frac{1}{g^2_0}+\frac{3h_G^\vee}{8\pi^2}\mathrm{log}\left( \frac{\mu}{M_{KK}}\right).
\end{equation}
Here $h_G^\vee$ is the dual Coxeter number of $G$ and for the rest of this section we will simply assume $G=SU(N)$ and one can just replace $h_G^\vee$ with $N$ throughout otherwise. Such a running relates the KK-scale to the confinement scale $\Lambda$ as
\begin{equation}
    \Lambda^3=\exp\left( \frac{-8\pi^2 }{g^2_0 N}\right)M^3_{KK}.
\end{equation}
Since $g^2\sim V_b^{-1}$, the RG flow to lower energies can be viewed as a shrinking of the $S^3_b$ volume to zero and seemingly to negative values below the confinement scale. From the duality of \cite{Atiyah:2000zz}, this suggests that the IR confined phase is captured by $M$-theory on $\Sigma (S^3_a/\Gamma_{\mathfrak{g}})$ for some fixed volume $V_{a,*}$ (the precise value depends on initial conditions of the RG flow as we highlight below). Indeed, $M$-theory on $\Sigma (S^3_a/\Gamma_{\mathfrak{g}})$ captures several qualities of the behavior of the confining phase of $\mathcal{N}=1$ SYM. This includes the fact there exist BPS domain walls, which correspond to $M5$-branes wrapping $S^3_a/\Gamma_{\mathfrak{g}}$, as well as strings with $Z(G)^{(1)}$ 1-form symmetry charge, which correspond to $M2$-branes wrapping minimal volume 1-cycles in the homology
\begin{equation}\label{eq:s3gammahomology}
    H_1(\Sigma (S^3_a/\Gamma_{\mathfrak{g}}))=Z(G)=\mathrm{Ab}(\Gamma_\mathfrak{g}).
\end{equation}
Here $\mathrm{Ab}(\Gamma_\mathfrak{g})$ denotes the abelianization of $\Gamma_\mathfrak{g}$. Additionally, one can see that $M$-theory on this background has a mass gap since the there are no $L^2$-normalizable harmonic forms \cite{Hausel:2002xg, Acharya:2024bnt} which would be required for dynamical massless degrees of freedom.

\begin{figure}[h]
    \centering
    \includegraphics[width=8cm]{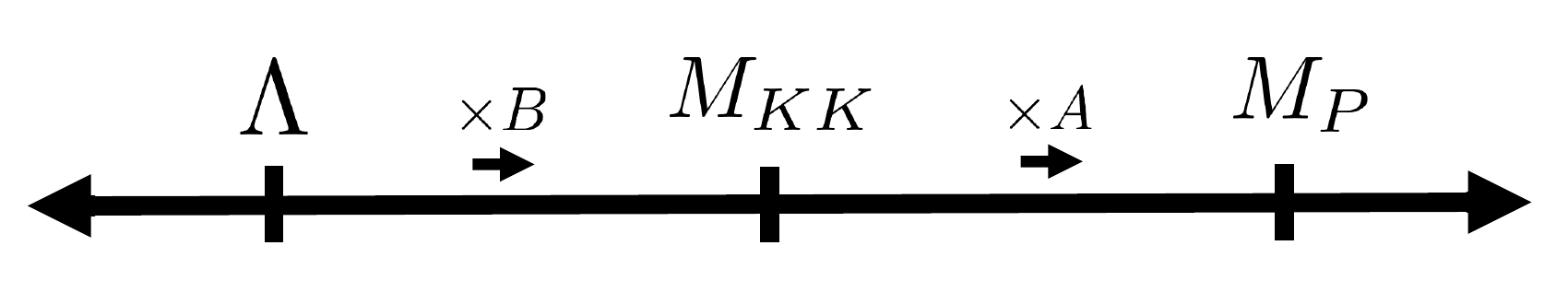}
    \caption{}
    \label{fig:scales}
\end{figure}

Solving for $V_{a,*}$ is not straightforward in general due to Euclidean $M2$-brane corrections. However, if we first assume that $V_{a,*}\geq M^{-3}_P$, then such corrections are parametrically small and we can trust the supergravity description of the geometry. In this regime we can match the BPS domain wall tension from compactifying the $M5$-brane on $S^3_a$
\begin{equation}\label{eq:wall}
    T_{wall}=T_{M5}\cdot V_{a,*}=M^{6}_P V_{a,*},
\end{equation}
with the fact that in a large $N$ limit with fixed 't Hooft coupling $\lambda_g\equiv g^2_0 N$, the $\mathcal{N}=1$ SYM domain wall tension is known to be
\begin{equation}
    T_{wall}=\frac{1}{2\pi^2}\langle \mathrm{Tr} \lambda \lambda \rangle = N\Lambda^3.
\end{equation}
The first equality follows from the exact result of \cite{Dvali:1996xe}, while the second follows from the fact that the gaugino condensate is proportional to $g^{-2}_0 \exp (-8\pi^2/Ng_0)$ \cite{Shifman:1987ia} while $g^{-2}_0\sim N$ in the 't Hooft limit. Altogether this allows us to express $V_{a,*}$ in terms of the 't Hooft coupling as
\begin{equation}
    V_{a,*}M^3_{P}=\lambda_g \exp\left( \frac{-8\pi^2}{\lambda_g}\right)
\end{equation}
which means that $V_{a,*}\geq M^{-3}_P$ requires large 't Hooft coupling. Going to such regime depends non-trivially on the three energy scales in this system, see Figure \ref{fig:scales}. Defining $A\equiv M_P/M_{KK}$ and $B\equiv M_{KK}/\Lambda$, we obtain a useful relation between these energy scales and the value of the base volume:
\begin{equation}
    V_{a,*}M^3_P=\frac{N}{(AB)^3}=\frac{N}{A^3}\exp\left(\frac{-8\pi^2 A^3}{N} \right).
\end{equation}
While decoupling gravity from the gauge theory physics amounts to taking $A \rightarrow  \infty$, there are two regimes of interest in taking this limit which will play a role in the rest of this note.

\paragraph{\textbf{MQCD Regime}}
At large $N$, if consider a limit $A\rightarrow \infty$ such that $A\sim N^{\alpha}$ where $\alpha<1/3$ then we see from above that
\begin{equation}
    B\sim \exp\left(N^{3\alpha-1}\right)\sim 1, \quad V_{a,*}M_P^3\sim N^{1-3\alpha}.
\end{equation}
This means that the volume of the $S^3_a/\Gamma$ base is large in Planck units and we can trust the $M$-theory geometric description of the low-energy physics. This regime is where we will be able to make the most precise statements about the confining string worldsheet. On the other hand, because $B\rightarrow 1$ we see that 4D field theory is Yang-Mills with additional massive adjoint chiral and massive (i.e. Higgsed) vector multiplets corresponding to the 7D KK-tower which is of order of the confinement scale. Such gauge theories arise commonly in stringy constructions (see for instance \cite{Maldacena:2000yy, Witten:1997ep, Hanany:1997hr}), and are called ``MQCD" to emphasize this technicality. The fact that the vacuum structure, presence of strings charged under $Z(G)^{(1)}$, and domain walls with correct IR behavior \cite{Acharya:2001dz}, match between this MQCD regime, using $M$-theory, and $\mathcal{N}=1$ QCD, using field theory, provides very strong evidence that these two field theories are identical in the deep IR, i.e. they are in the same universality class. Finally, notice that from applying \eqref{eq:gammalength} at a arbitrary radial coordinate $r$, we see the length of the torsion 1-cycle is below the Planck scale for $r\in [r_a,N]$ and considering that $r_a\sim N^{\frac{2-3\alpha}{3}}$, there is a large region around the zero-section which is equivalent to IIA string theory on the resolved conifold with $N$ units of RR-flux on the base $\mathbb{P}^1$ (A-type cases) or $\mathbb{RP}^2$ (D-type cases).

\paragraph{\textbf{SYM Regime}}
If we instead consider the large-$N$ behavior $A=(CN)^{1/3}$ for some fixed constant $C> 1$, then we have that
\begin{equation}\label{eq:qcdv}
    B\sim e^C, \quad V_{a,*}M_P^3\sim \frac{e^{-C}}{C} N^{0}.
\end{equation}
The limit where we decouple the 7D KK-modes amounts to taking the $C\gg 1$ which means that this $M$-theory geometry engineers 4D $\mathcal{N}=1$ SYM. The tradeoff is that now the volume of the base $S^3_a/\Gamma$ is below the Planck scale \footnote{Note that because the relation \eqref{eq:qcdv} relied on \eqref{eq:wall} which is strictly speaking only valid at large volume, one can use a use a similar formula for using the $D4$-brane tension in the IIA picture to arrive at a similar dependence on $C$.} where the description of the physics in the region $r_a\leq r<M^{-1}_P$ still admits a IIA string theory description.


\section{$M$-theory vs. Field Theory Parities}\label{sec:versus}
The key reason why there is a difference between what an $M$-theorist would call spatial reflection versus what a 4D gauge theorist studying the engineered $\mathcal{N}=1$ (M)QCD theory would call it boils down to the fact that the $M$-theory 3-form potential $C_3$ is odd under orientation reversal \footnote{For a $p$-form $A$ to be even under $\mathsf{R}_1$ means, in components, that $A_{\mu_1...\mu_p}$ is even if no components are along the 1-direction, while $A_{1\mu_2...\mu_p}$ is odd. These conditions are reversed if $A$ is odd under $\mathsf{R}^{11D}_1$ and these are sometimes called pseudo-$p$-forms.}:
\begin{equation}
    \mathsf{R}^{11D}_1: C_3\rightarrow -C_3.
\end{equation}
This property is fixed by the requirement that the 11D Chern-Simons term $(1/6)\int C_3G_4G_4$ is invariant, and this means the maximal torus gauge fields $A^{a=1,...,\mathrm{rank}(\mathfrak{g})}$ as obtained in the $G_2$-geometry by integrating over collapsed exceptional cycles transform as:
\begin{equation}\label{eq:aa}
  \mathsf{R}^{11D}_1: A^a\equiv \int_{E_a}C_3 \; \rightarrow\;   -A^a.
\end{equation}
If we regard $A^a$ as abelian gauge fields, then the action of $\mathsf{R}^{11D}_1$ appears as $\mathsf{C}\mathsf{R}$ to usual field theory conventions where the vector potential and electric fields are even while magnetic fields are odd under $\mathsf{R}_1$. Equivalently, electric charges under these gauge fields flip sign under $\mathsf{R}^{11D}_1$ which is what is normally associated with $\mathsf{C}\mathsf{R}_1$ \footnote{The converse is true of magnetic charges which is consistent with the $M$-theory statement that $C_6$ is even under $\mathsf{R}^{11D}_1$.}. Generalizing these statements to full non-abelian group is more subtle since $\int_{E_a}C_3$ is not a $\mathfrak{g}$-gauge invariant expression, however it is a gauge invariant statement whether \eqref{eq:aa} is an outer or inner automorphism of $\mathfrak{g}$. The latter/former case means that \eqref{eq:aa} can/cannot be undone by a $\mathfrak{g}$-gauge transformation. Given that an infinitesimal gauge transformation in the maximal torus directions is generated by $iA^a\mathfrak{t}^a$ we can regard \eqref{eq:aa} as a sign flip of the Cartan generators of $\mathfrak{g}$, this means we can rewrite \eqref{eq:aa} as
\begin{align*}
    &\textnormal{\textit{Given a vector in the weight lattice of $G$, $w\in \Lambda^{\mathrm{wt}}_{G}$,}}\\
    &\textnormal{\textit{then we have that} }\mathsf{R}^{11D}_{1}(w)=-w.
\end{align*}
Now, recall that there is a short exact sequence of groups
\begin{equation}\label{eq:latticeseq}
  0\rightarrow \Lambda^{\mathrm{rt}}_{G}\rightarrow \Lambda^{\mathrm{wt}}_{G}\rightarrow Z(G)\rightarrow 0
\end{equation}
where $\Lambda^{\mathrm{rt}}_{G}$ is the root lattice of $G$. This means that the action of $\mathsf{R}^{11D}_1$ descends to $Z(G)$ as just a multiplication by a sign. Since non-trivial outer automorphisms of $G$ always descend to non-trivial outer automorphisms of $Z(G)$, and the multiplications by signs only affect elements that are not of order-2 we find the necessary and sufficient condition for \eqref{eq:aa} to be an outer automorphism is
\begin{align*}
    &\textnormal{\textit{$\mathfrak{g}$ must be such that the center of its uniquely }} \\
    &\textnormal{\textit{associated simply connected group, $Z(G)$, has a}}\\
     &\textnormal{\textit{non-trivial element which is not of order-2.}}
\end{align*}
An equivalent way of stating this is that the background field for the 1-form symmetry, under which the QCD string is charged, transforms as
\begin{equation}
    \mathsf{R}^{11D}_{1}:\;  b_2\equiv \int_{\gamma_1} C_3 \rightarrow -b_2
\end{equation}
and since this background field is valued $Z(G)$ as one is integrating over a generating 1-cycle $\gamma_1\in H_1(S^3_a/\Gamma_{\mathfrak{g}})=Z(G)$, $-b_2\neq b_2$ only when the generators of $Z(G)$ are not of order-2. From the list of gauge group centers:
\begin{align*}
    & Z(SU(N))=\mathbb{Z}_N, \;  Z(Spin(4N))=\mathbb{Z}^2_2, \\
    & \; Z(Spin(4N+2))=\mathbb{Z}_4, \;  Z(E_6)=\mathbb{Z}_3 \\
    & Z(Spin(2N+1))=Z(Sp(N))=Z(E_7)=\mathbb{Z}_2
\end{align*}
we see that the gauge algebras which satisfy $\mathsf{R}^{4D}_1=\mathsf{R}^{11D}_{1}$ are
\begin{equation}\label{eq:summary1}
\mathfrak{g}=\mathfrak{so}(4N), \; \mathfrak{sp}(N), \; \mathfrak{so}(2N+1), \; \mathfrak{e}_7.
\end{equation}
We listed non-simply laced algebras above which can be constructed from variations of the Bryant-Salamon constructions we have considered above albeit with either frozen singularities (i.e. turning on a background $ \int_{S^3_a/\Gamma}C_3=\frac{1}{2}$ \cite{deBoer:2001wca, Atiyah:2001qf}) or with the more general quotients recently studied in \cite{Khlaif:2025jnx}. Meanwhile those 4D $\mathcal{N}=1$ SYM theories which have $\mathsf{R}^{11D}_{1}$ acting by an outer automorphism on the gauge algebra and thus require a internal isometry $\mathsf{R}_{int}$ which appears as charge conjugation so that one can define what a 4D field theorist would regard as reflection $\mathsf{R}^{4D}_1=\mathsf{R}^{11D}_{1}\circ \mathsf{R}_{int}$ have gauge algebras:
\begin{equation}\label{eq:summary2}
  \mathfrak{g}=\mathfrak{su}(N>2), \; \mathfrak{so}(4N+2), \; \mathfrak{e}_6
\end{equation}

Closing this section with an example, let us take $\mathfrak{g}=\mathfrak{e}_6$. The highest weight of the $\mathbf{27}$ representation can be presented as $(1,0,0,0,0,0)$ while its lowest weight as $(0,0,0,0,-1,0)$ \cite{Slansky:1981yr}. Meanwhile, the $\overline{\mathbf{27}}$ representation has $(0,0,0,0,1,0)$ and $(-1,0,0,0,0,0)$ as its highest and lowest weight respectively. Therefore $\mathsf{R}^{11D}_{1}$ exchanges $\mathbf{27}$ with $\overline{\mathbf{27}}$. To emphasize that the conclusions of this section are irrespective of the choice of gauge group topology, for both $E_6$ or $E_6/\mathbb{Z}_3$ (S)YM theory, there are stable QCD strings at low-energy which are created by Wilson lines in the $\mathbf{27}$ representation. The only difference is that in the $E_6/\mathbb{Z}_3$ case, such an operator must be the boundary of a topological surface operator. This is reflective of the fact that the QCD string is charged under a $\mathbb{Z}_3$ 1-form gauge symmetry in this case.


\section{Charge Conjugation Isometries of Bryant-Salamon Orbifolds}\label{sec:ccisometries}

We now state the geometric conditions required for an isometry $\mathsf{R}_{int}$ to appear as a charge conjugation in the 4D gauge theory. We will first focus on the A-type cases before commenting on the D- and E-type cases. To fit the bill, $\mathsf{R}_{int}\in G_{isom.}(\Sigma(S_a^3/\mathbb{Z}_N))$ must satisfy:
\begin{align*}\label{eq:rintconditions}
  \mathrm{1.}& \quad \textnormal{$\mathsf{R}^2_{int}=1$}\\
  \mathrm{2.}& \quad \textnormal{\textit{$\mathsf{R}_{int}$ reverses the orientation of $\gamma_1$. This implies that}} \\
 & \quad \textnormal{\textit{the uplift of $\mathsf{R}_{int}$ to $\widetilde{\mathsf{R}}_{int}\in G_{isom.}(\Sigma S^3_a)$ satisfies:}} \\
 & \quad \quad \quad  \quad \quad \quad  \quad  \quad \widetilde{\mathsf{R}}^{-1}_{int}u_0\widetilde{\mathsf{R}}_{int}=u^{-1}_0
\end{align*}
The first condition simply reproduces the field theory relation \footnote{One is always free to take the convention where charge conjugation squares to fermion parity $\mathsf{C}^2=(-1)^F$ but this does not matter for our purposes since we are focused on its action on bosonic degrees of freedom.} $\mathsf{C}^2=1$. The second condition can be argued for as follows. First, notice that such a $\mathsf{R}_{int}$ acts as a $\mathbb{Z}_2$ outer automorphism on the first homology of the boundary $H_1((S^3_a/\mathbb{Z}_N)\times S^3_b,\mathbb{Z})=\mathbb{Z}_N=\langle [\gamma_1] \rangle$ which flips $[\gamma_1]\rightarrow -[\gamma_1]$ (assuming $N>2$). Since the Wilson and 't Hooft line operators for the 4D gauge theory can be built by wrapping M2 and M5 branes on $\mathrm{Cone}(\gamma_1)$ and $\mathrm{Cone}(\gamma_1)\times S^3_{b}$ respectively \cite{Albertini:2020mdx}, we see that $\mathsf{R}_{int}$ flips the electric and magnetic charges as expected from charge conjugation.

We can reach this same conclusion in another way if we zoom in on a local patch of the zero-section $U_{loc.}\subset S^3_b\subset \Sigma S^3_b$, the $G_2$-manifold resembles $U_{loc.}\times \mathbb{C}^2/\mathbb{Z}_N$ and if we resolve the second factor to a Gibbons-Hawking space \footnote{This procedure costs some energy since the Coulomb branch of the 7D SYM theory is lifted due to the compactification, but the point being made here is a topological one.}, $\tilde{X}_N$, we have the short exact sequence in relative homology
\begin{equation}\label{eq:mvseq}
  0\rightarrow H_2(\tilde{X}_N)\rightarrow H_2(\tilde{X}_N, \partial \tilde{X}_N)\rightarrow H_1(\partial \tilde{X}_N)\rightarrow 0
\end{equation}
which equates the homology class $[\gamma_1]$ which fractional linear combinations of the exceptional 2-cycles generating $H_2(\tilde{X}_N)$. Indeed this short exact sequence matches that of \eqref{eq:latticeseq}. The $\mathbb{Z}_2$ action on $[\gamma_1]$ then descends to the exceptional 2-cycles which makes $\int_{E_a} C_3$ odd under $\mathsf{R}_{int}$. We can conclude then that a $\mathsf{R}_{int}$ satisfying the two conditions above, together with \eqref{eq:4d11drefl} defines the correct notion of reflection transverse to the QCD string $\mathsf{R}^{4D}_1$.


We now enumerate the possible choices of $\mathsf{R}_{int}$. We find it convenient to present these actions in quarternion coordinates where $u=z_{1u}+jz_{2u}\in SU(2)_u$ and similarly for $v\in SU(2)_v$ and $w\in SU(2)_w$. The actions on $a$, $b$, and $c$ in \eqref{eq:isometryactions} can just be regarded a quarternion multiplication where $a=(z_{1a}+jz_{2a})/||z_{1a}+jz_{2a}||$ and similarly for $b$ and $c$. When $N$ is odd, one can check by exhaustion that the only possible choices of $\mathsf{R}_{int}\in G_{isom.}(\Sigma(S^3_a/\mathbb{Z}_N))$ satisfying the three conditions above are (up to independent choices of signs for $u$, $v$, and $w$)
\begin{equation}\label{eq:rintnodd}
  \mathsf{R}_{int}= (u,v,w) = (j \; \mathrm{or} \; k, i \; \mathrm{or} \; j \; \mathrm{or} \; k, i \; \mathrm{or} \; j \; \mathrm{or}\;  k  ).
\end{equation}
Each choice of $u$, $v$ and $w$ is independent in the above so there are a total of $18$ possible choices of $\mathsf{R}_{int}$ up to sign choices. To illustrate why these choices are exhaustive, let us for instance take $\mathsf{R}_{int}=(u,v,w)=(j,i,k)$. In this case we indeed have an action that squares to $1$ since $\mathsf{R}_{int}^2=(-1,-1,-1)$ acts trivially on $(a,b,c)$. Furthermore we have that $j^{-1}u_0 j=-j\exp(2\pi i /N) j=-\exp(-2\pi i /N) j^2=\exp(-2\pi i /N)=u_0^{-1}$ which means that $[\gamma_1]\rightarrow -[\gamma_1]$. Notice that a choice of $u=i$ does not work because it commutes with $u_0$. When $N>2$ even, the additional possibilities are
\begin{equation}\label{eq:noddrint}
  \mathsf{R}_{int}=(u,v,w)= (j,1,1), \; \mathrm{or} \; (k,1,1)
\end{equation}
because now $\mathsf{R}_{int}^2=(-1,1,1)=(u^{N/2}_0,1,1)\sim (1,1,1)$. When $N=2$ there is no such action simply because there is no outer-automorphism on the homology group of 1-cycles $H_1(\Sigma(S^3_a/\mathbb{Z}_2))=\mathbb{Z}_2$, making the second condition above impossible to satisfy.

\paragraph{\textbf{Other Gauge Algebras}}
Now focusing on the orbifold groups $\Gamma_{D_{2N+1}}$ and $\Gamma_{E_6}$, as reviewed in Section \ref{sec:review}, the commutator subgroup of these $\Gamma$s with $SU(2)_u$ is $\mathbb{Z}_2$ which acts as an outer automorphism of $\Gamma$. Let us generally call this element $\rho\in SU(2)_u$ which is given in \eqref{eq:elementrho}. Then we see that the possible choices of $\mathsf{R}_{int}$ are of course $(\rho, 1,1)$, as well as
\begin{equation}
    (\rho, i, i), \; (\rho, j, j), \; (\rho, k, k)
\end{equation}
which square to the identity because the $\mathrm{diag}(-1,-1)$ is an element for these $\Gamma$. Note that even though there is an analogous $\rho$ for the $D_{2N}$, because $\mathsf{R}^{11D}_1=\mathsf{R}^{4D}_1$ in these cases, such symmetries are independent of the reflection symmetries.



\section{QCD String Worldsheet Physics}\label{sec:QCDstringworldsheet}

We now discuss the 2D worldsheet theories for the QCD strings as obtained from reducing M2-branes on the minimum length torsional 1-cycle $\gamma_1$ in both the MQCD and SYM regimes. First addressing the tension of these strings, consider the large-$N$ scaling $A\sim N^{\alpha}$ in the notation of Section \ref{sec:g2flopconfinement}. We saw that this implies that $V_{a,*}\sim N^{1-3\alpha}$ in 11D Planck units which implies that the parameter $r_a$ (the radius of $S^3_a$ before orbifolding) scales as $r^3_a\sim N^{2-3\alpha}$. From \eqref{eq:gammalength}, we see then see that the (M)QCD string tension scales as
\begin{equation}
    T_{string}\sim N^{-\alpha-\frac{1}{3}}M_P^2\sim N^{\alpha-\frac{1}{3}}\Lambda^2.
\end{equation}
In the SYM regime, we have $\alpha=1/3$ which matches large-$N$ lattice predictions that the QCD string scales as $T_{string}\sim N^0$ (for a review see \cite{Teper:2002kh}). Meanwhile taking $0<\alpha<1/3$ specifies an MQCD which means that its string tension differs from the SYM prediction at leading order \footnote{In \cite{Witten:1997ep}, a different string construction of 4D $\mathcal{N}=1$ gauge theories was given which also yields an MQCD-like theory with KK-masses of order of the confinement scale. It was suggested there that $T_{string}\sim N^0$ for their MQCD strings but this is not in contradiction with our result for two reason. The first is that this reference imposes this scaling by hand, and moreover the MQCD theory in \cite{Witten:1997ep} is different than the MQCD theory in this note (different dimension of KK-tower etc.). }.

\subsection{MQCD Regime}
In the MQCD setting, the $M$-theory $G_2$-geometry can be trusted as a description of the M2-brane target space physics because the zero-section volume is far above the Planck scale. As we saw above however, the length of the 1-cycle on which we wrap an M2-brane to obtain the MQCD string is smaller than the Planck length, so one may wonder whether one can obtain the MQCD worldsheet theory just from dimensionally reducing the low-energy M2-brane effective action. One can see that this is a valid step simply because the reduction of the Green-Schwarz sigma model action of the M2-brane on this $S^1$ is known to reduce to the Green-Schwarz sigma model of the IIA fundamental string \cite{Duff:1987bx, Townsend:1995kk}, and furthermore the curvature of the directions normal to the $S^1$ is small so such a reduction can be performed adiabatically.

We now present the low-energy NLSM action for a static M2-brane placed on $\gamma_1$ in the $G_2$-manifold, and along the $X_3$-direction in $\mathbb{R}^{3,1}$. The normal coordinates include $X_1$ and $X_2$, the radial direction in the Bryan-Salamon space $r$, while the four other normal coordinates can be presented as six complex coordinates $z_{iJ}$, $i=1,2$, $J\in \{a,b,c\}$, subject to the constraints
\begin{align}\label{eq:constraints}
  |z_{1J}|^2+|z_{2J}|^2=f_{J}(r,r_a), \quad  \prod_{J\in \{a,b,c\}}\begin{pmatrix}
            z_{1J} & -z^*_{2J} \\
      z_{2J} & z^*_{1J}
           \end{pmatrix}=\mathbf{1}_{\mathrm{2x2}}
\end{align}
where the functions $f_J$ can be read off from \eqref{eq:sigmas3metric} to be $f_a=\frac{1}{36}(r^2+(\frac{r^3_a}{2r}))$, and $f_b=f_c=\frac{1}{36}(r^2+(\frac{r^3_a}{r}))$. The Nambu-Goto (NG) action for the M2-brane in the static gauge takes the general form
\begin{align*}
    -T_{M2}\int_{M2} \bigg[-\mathrm{det}\bigg( \eta_{\alpha \beta}+c^2\big(\frac{1}{2}\partial_\alpha X^{i} \partial_\beta X^{i}+\frac{1}{2}g_{rr}(r)\partial_\alpha r \partial_\beta r\\+
    \frac{1}{2}f_J(r)\partial_\alpha z_{iJ}\partial_\beta z^*_{iJ} -2\overline{\lambda}\gamma_\alpha\partial_\beta \lambda\big)+c^3(\mathrm{4 \;Fermi \; term})\bigg)
    \bigg]^{1/2}
\end{align*}
where $c=M^{-2}_P$ in the convention where the bosonic fields have mass dimension one. We argue that in the $G_2$ geometric directions normal to $\gamma_1$, we can restrict to the $r=r_a$ locus, i.e. just two coordinates in $S^3/\Gamma_\mathfrak{g}$ normal to $\gamma_1$. This is because the $S^{3}/\Gamma_\mathfrak{g}$ volume at $r=r_a$ is at a minimum (see \eqref{eq:sigmas3metric}) so any deformations of the $M2$-brane along $\gamma_1$ in the $r$-direction will necessarily increase its length. This intuitively implies that any excitations in this radial direction will have a mass proportional to the scale of the $M2$-brane tension, which is $M_P$. In fact after expanding the square root in the NG action, the equations of motion for $r$ with constant 2D mass $\sim k_r N^{1/3 +\alpha} M_P$ becomes
\begin{equation}\label{eq:reom}
  (r^4-(r^3_a)r)\partial^2_\theta r +3(r^3_a)(\partial_\theta r)^2+2(r^3_a)k^2_r r^2+k^2_r r^5=0
\end{equation}
where $\theta$ is the coordinate along $\gamma_1$, and we have taken the fields parameterizing $S^3_b$ to be constant without much loss of generality \footnote{Since the kinetic terms for such fields are proportional to a function of $r$ which vanishes when $r=r_a$, this approximation of \eqref{eq:reom} is same for small $r-r_a$}.
We find that there is no classically 2D massless ($k_r=0$) solutions to \eqref{eq:reom} which are periodic along $\theta$ and any solution with $k_r\neq 0$ would lead to a classical 2D mode with mass greater than the Planck scale. The low-energy spectrum of the MQCD string then consists of two massless spin-0 bosons $X^{i=1,2}$ (goldstone bosons for breaking $\mathbb{R}^2_{1,2}$ translation symmetry), a massless Dirac fermion (goldstino for breaking 4 supercharges), and two classically light spin-0 bosons parametrizing in the normal directions of $\gamma_1$ in $S^3_a/\Gamma_{\mathfrak{g}}$. These massless excitations are apriori expected for any non-BPS string in a 4D $\mathcal{N}=1$ system. On the other hand, the latter two are not protected from any symmetry due to quantum corrections, so their mass is pushed up to the scale $\Lambda$. We now turn to studying the reflection quantum numbers of these two massive spin-0 bosons before discussing their quantum correction in more detail.


\paragraph{\textbf{Worldsheet Pseudoscalars}}
We now test the reflection quantum numbers for the two classically light scalars. As we saw for the gauge algebras listed in \eqref{eq:summary1} all of these modes are parity-even because there is no composition with some internal isometry $\mathsf{R}_{int}$. Meanwhile for the gauge algebras listed in \eqref{eq:summary2}, $\mathsf{R}^{4D}_1$ does include the action of an internal isometry and from Section \ref{sec:ccisometries} we saw that there were a plurality of ways to define $\mathsf{R}_{int}$. Most of these will not preserve the $\gamma_1$ locus. Restricting to the A-type cases for concreteness, the $\gamma_1$ locus is located at $r-r_a=z_{2a}=0$, so we see for instance $\mathsf{R}_{int}=(j,i,i)$ is not a valid choice to test the quantum numbers of the two light worldsheet scalars because
\begin{equation}\label{eq:badexample}
  j(z_{1a}+jz_{2a})i^{-1}=kz_{1a}-iz_{2a}
\end{equation}
does not preserve the condition that the $j$ and $k$ components vanish at $\gamma_1$. One finds that the valid choices of $\mathsf{R}_{int}$ reduce to six (up to signs)
\begin{equation}\label{eq:validrint}
  \mathsf{R}_{int}= (j,j,i \; \mathrm{or}\; j \; \mathrm{or} \; k), \; (k,k,i \; \mathrm{or}\; j \; \mathrm{or} \; k).
\end{equation}
In terms of coordinates of the normal space $S_a^2$ parametrized by $r_{1a}+jz_{2a}$ satisfying
\begin{equation}\label{eq:s2zn}
  S_a^2: \; r^2_{1a}+|z_{2a}|^2=f_a,
\end{equation}
these three classes of $\mathsf{R}_{int}$ act as
\begin{align}\label{eq:3transformations}
  u=v=j: \; r_{1a}+jz_{2a} \; \rightarrow & \; r_{1a}+jz^*_{2a} \\
  u=v=k: \; r_{1a}+jz_{2a} \; \rightarrow  &\;  r_{1a}-jz^*_{2a}.
\end{align}
These also specify $SU(2)$ rotations acting on $S^3_b$, and we see from \eqref{eq:3transformations} that all of these choices for $\mathsf{R}_{int}$ imply that there are is a pseudoscalar axion $\theta_{2a}=\mathrm{arg}(z_{2a})$ under $\mathsf{R}^{4D}_1$. One can also make this explicit in the D- and E-type cases, but there is a simple conceptual reason why there must always be a pseudoscalar for each of the gauge algebras in \eqref{eq:summary2}. This is because $\mathsf{R}_{int}$ acts a rotation on $S^3_a/\Gamma_{\mathfrak{g}}$ while sending $[\gamma_1]\mapsto -[\gamma_1]$ and because rotations preserve orientation, this means that $\mathsf{R}_{int}$ must simultaneously flip the normal bundle of $\gamma_1$ inside $S^3_a/\Gamma_{\mathfrak{g}}$ which implies that one of the light spin-0 bosons is a scalar and the other is a pseudoscalar. We will call these $\phi$ and $a$ below respectively.

\paragraph{\textbf{Quantum Corrections}}
As mentioned above, there is no symmetry argument for why the classical masses of $a$ and $\phi$ should be protected from being of order of the confinement scale. For the A-type case, we can be a bit more concrete as the M2-worldvolume has a target space direction $S^2_a$ parametrized by $a=\theta_{2a}$ and $\phi=|z_{2a}|$. Upon compactifying on $S^1_\gamma$, $\phi$ is periodic while $a(\theta_{1a}+L_\gamma)=a-\frac{2\pi}{N}$. This means that $a$ has a classical mass of order $(NL_{\gamma})^{-1}$ in Planck units which scales as $N^{-2/3}\Lambda$ so is effectively classical massless and is minor compared to the possible quantum corrected mass. This $S^2_a$ NLSM sector then effectively reduces to the 2D $O(3)$ model. This theory is expected to confine with a mass gap calculated in \cite{Hasenfratz:1990zz} to be an to be an order one constant times $\Lambda$. This would receive only minor corrections for the MQCD string due to the small classical mass mentioned above, and irrelevant interactions with $X^i$ and $\lambda$. One such interesting interaction term that will appear in the effective action is
\begin{equation}\label{eq:topinv}
  c\Lambda^{-2}\int_{\mathbb{R}^{1,1}} a\epsilon^{ij}\epsilon^{\alpha\beta}\partial_\alpha \partial_\gamma X^i\partial_\beta\partial^\gamma X^j
\end{equation}
where $c$ is a to-be-determined constant. The integral $\int\epsilon^{ij}\epsilon^{\alpha\beta}\partial_\alpha \partial_\gamma X^i\partial_\beta\partial^\gamma X^j$ is a topological invariant which counts the signed self-intersections of the worldsheet in the 4D spacetime. Such a term is technically natural as intersection numbers flip sign under $\mathsf{R}^{4D}_{1}$, the effective action term \eqref{eq:topinv}.

\subsection{SYM Regime}
In this regime, the zero-section of the $G_2$-geometry is below the Planck scale which complicates the M2-brane sigma model picture in the MQCD case. As pointed out above however, placing the string at a radial coordinate $r<M^{-1}_{P}$ reduces the worldsheet theory to IIA string on a conifold in the A-type case or to an orientifold of the conifold in the D-type case with RR-flux. One can reason that the M2-brane will stay in this region because if one places an M2-brane on $\gamma_1$ but at some radial coordinate $r>M^{-1}_{P}$, then the low-energy NG action will evolve the M2-brane to a smaller radius and thus into the IIA region. From the lattice result that $T_{string}\sim N^0$, it appears that the M2-brane still favors the minimal radius of the $G_2$ manifold, and in the IIA language, the coupling constant is of the order $g_s=L_\gamma\sim e^{-C}/(NC)$ which means that the stringy 't Hooft coupling is small $g_sN\sim e^{-C}/C \ll 1$ since we take the limit $C\rightarrow \infty$. The significance of this is that this means that the effective of the RR-flux on the IIA worldsheet, which can be seen as a non-local deformation in the RNS language \cite{Berenstein:1999jq, Berenstein:1999ip, Cho:2018nfn}, small and thus well approximated by the conifold geometry. Since the geometric analysis of $\mathsf{R}_{int}$ just descends to the conifold geometry for the A- and D-type cases, we can again conclude the same pseudoscalar pattern as for MQCD strings. For the E-type cases, we still have that a stable M2-brane wrapping $\gamma_1$ will be in the region $r<M^{-1}_{P}$ and that there will be some pseudoscalar operators at low-energies in the $E_6$ case but we cannot yet be completely sure if this remains of the order of the confinement scale or is sent far above it and decouples.

\section{Non-Supersymmetric Yang-Mills from $G_2$-manifolds}\label{sec:nonsusy}

In this section we describe non-supersymmetric four dimensional theories with a mass gap and confinement.
Recall that one can deform $\mathcal{N}=1$ SYM to non-susy YM by a relevant deformation, i.e. by giving the gaugino an infinite mass. From the point of view of the string worldsheets, this deformation is necessarily relevant (for instance the 2D massless Dirac goldstino will gain a mass proportional to the gaugino mass) and although we do not know how to calculate the corrected masses of $a$ and $\phi$ which could be so large that they are decoupled from the system, we can be sure that there are no new excitations after the deformation. We can also be sure that the quantum numbers of the massive spin-0 excitations won't change under this 2D RG flow so long as there are no dangerously irrelevant operators \footnote{A dangerously irrelevant operator whose dimension-full coupling goes to infinity as the mass of the goldstino goes to infinity could cause the formation of a finite mass bound state between the spin-0 boson and the pseudoscalar bilinear of the goldstino. We find this scenario unlikely but cannot rigorously rule this out.}. Of course lattice evidence strongly suggests this doesn't happen for $a$ in the $\mathfrak{su}(N>2)$ cases. However, we \textit{can} conclude that
that the list of gauge algebras with no pseudoscalars in \eqref{eq:summary1} will continue to not have pseudoscalars.
It would be very interesting to investigate these ideas in lattice gauge theory.

There is a simple way to break supersymmetry within the framework of $M$-theory on $\Sigma S^3/\Gamma$ with the $G_2$-holonomy metrics we have discussed. Since the smooth manifolds  $\Sigma S^3/\Gamma$ are non-simply connected they will in general admit more than one choice of spin structure. In the rest of this paper we have been implicitly assuming, due to the fact that the metric has $G_2$-holonomy, a canonical choice of spin structure: one for which there exists a covariantly constant i.e. parallel spinor with respect to the Levi-Cevita connection. However, if $\Gamma$ contains elements of even order there could be multiple choices of spin structure and only one of them gives rise to a supersymmetric spacetime. E.g. $\Gamma \cong \mathbb{Z}_{2k}$ has two choices of spin structure, only one of which is supersymmetric.


Such a Scherk-Schwarz-like spin structure exists whenever $\mathrm{Hom}(\pi_1(\Sigma (S^3_a/\Gamma_{\mathfrak{g}})),\mathbb{Z}_2)=H^1(\Sigma (S^3_a/\Gamma_{\mathfrak{g}}),\mathbb{Z}_2)\neq 0$ which is the case when $\mathrm{Ab}(\Gamma_{\mathfrak{g}})(=Z(G))$ contains an order-2 element. Such a background is non-supersymmetric, has $\mathrm{Ab}(\Gamma_{\mathfrak{g}})$ 1-form charged string states, and has a semiclassically stable vacuum due to the fact that the supergravity equations of motion due to the vanishing of the Ricci tensor. Moreover, this vacuum has a mass gap because (just as in the supersymmetric version) there are no $L^2$-normalizable harmonic forms. Therefore, this is a smooth $M$-theory spacetime giving rise to a four-dimensional vacuum with all the properties expected of the vacuum of YM theories with gauge groups $SU(2k), SO(2k), E_7$.  The bosonic sector of the long confining strings discussed above would naively seem to give a reasonable semi-classical description of the strings in this non-supersymmetric confining vacuum.

\section{Conclusions/Outlook}

In this note, we have shown how the $M$-theory realization of 4D $\mathcal{N}=1$ Super-Yang-Mills can reproduce a signature first identified from the lattice. This feature concerns the necessity of including a massive pseudoscalar in the 2D EFT of the QCD string worldsheet (for 4D $\mathcal{N}=0$ $SU(N\geq 3)$ Yang-Mills) in order to match various Wilson line correlation functions. Our $M$-theory analysis moreover predicts that such a pseudoscalar should not be present for the gauge algebras
\begin{equation}
    \mathfrak{g}=\mathfrak{su}(2), \; \mathfrak{sp}(N), \; \mathfrak{so}(2N+1), \mathfrak{so}(4N), \mathfrak{e}_7.
\end{equation}
The details of our construction imply that this prediction is, strictly speaking, valid at large-$N$ and $\mathcal{N}=1$. We have also given arguments for why this plausibly holds for non-supersymmetric Yang-Mills as well.

We anticipate several ways in which our analysis of QCD string worldsheet theories, including the qualitative predictions on the presence of QCD string axions, can be generalized. For one, we have restricted ourselves to QCD strings of minimal 1-form charge, but more general charged strings (so-called k-strings) would involve bound states of M2-branes, and one can investigate the nature of pseudoscalars there. Another generalization is considering different global forms of the gauge groups. While we have shown this does not affect the presence/absence of QCD string axions, this can affect topological terms in the effective action which may affect the phase structure of QCD string. Similarly, for backgrounds with a frozen flux, there will be an additional theta angle present in the 2D effective action which weights Euclidean instantons associated with wrapping $S^3_a/\Gamma_{\mathfrak{g}}$ by M2 by a $\pi$-phase \footnote{This for instance will lead to a topological difference between the $\mathfrak{sp}(k)$ QCD string theories built from adjoining D-type BS orbifolds with a $C_3$ monodromy along the 3-cycle $[S^3/\Gamma_{D}]$ versus obtaining the $\mathfrak{sp}(k)$ through a more general quotient as in \cite{Khlaif:2025jnx} where a $\mathfrak{su}$ gauge factor is folded to $\mathfrak{sp}$ along a 1-cycle. The authors are unaware of such distinctions present in the field theory literature which would be an interesting direction to explore. }. Additionally, one can study QCD strings for 3D $\mathcal{N}=2$ SYM by simply studying IIA string theory on the same BS-orbifold. Each of these cases could have implications for non-susy Yang-Mills lattice studies \footnote{See for instance \cite{Dubovsky:2014fma} which claims the existence of massive \textit{scalars} on the worldvolumes of certain string bound states.}.

The general lesson highlighted in this note of carefully identifying how a field theory orientation reversal is lifted to its string construction, should lead to several more identifications of pseudoscalars in QCD string worldsheets, even those constructed from holography. For example, in the Klebanov-Strassler model, the confining 4D gauge theory is invariant under $\mathsf{R}^{4D}_i$ but the full IIB background is chiral. One must then compensate by a reflection in the direction normal to the D3/D5-brane stack so that $\mathsf{R}^{4D}_i$ is really a 10D rotation from the IIB perspective. See \cite{Gomis:2025gzb, Dierigl:2023jdp} for similar remarks. Since the confining string in this case is just an F1-string, then there will be necessarily be an odd-scalar associated with such a transformation.

Finally, a central goal in future QCD string studies from this $M$-theory would be to make our conclusions more qualitative, i.e. what are the precise values of the masses of $a$ and $\phi$? For A- and D-type cases, we saw that the large-N SYM regime is well-approximated by IIA on a RR-flux background, can recent advances in studying such backgrounds from the closed-string-field-theory perspective \cite{Cho:2018nfn} be of use? Additionally, it would be interesting to understand how the intuition mentioned in the Introduction that the QCD string axion arises due to the UV variable $\mathrm{Tr}\exp(i\oint A)F_{12}$, is modified for the gauge algebras for which we predict there to be no axion. Perhaps the UV variable $\mathrm{Tr}\exp(i\oint A) [\epsilon_{\mu\nu\rho \lambda}F^{\mu\nu}F^{\rho\lambda}]F_{12}$ can be matched in the IR to one of the propagating \textit{scalars} in these cases, as discussed in Section \ref{sec:QCDstringworldsheet}? It would also of course be exiting to study 4D $\mathcal{N}=1$ QCD strings directly from the lattice to more rigorously test our predictions from $M$-theory!




\begin{acknowledgements}
We thank Jan Albert, Andreas Athenodorou, Iosif Bena, Michele Del Zotto, Barak Gabai, Victor Gorbenko, Andrea Guerrieri, Manki Kim, Zohar Komargodski, Shota Komatsu, Joan Elias Mir\'o, Leonardo Rastelli, and Cumrun Vafa for helpful discussions/correspondence. We thank the organizers of the Strings and Geometry 2025 conference where this work was initiated. ET thanks the Aspen
Center for Physics, which is supported by National Science Foundation grant PHY-2210452,
for their hospitality during which part of this work was carried out. ET thanks the 2025 Simons Summer Workshop, IPhT Saclay, IHES, the University of Bern, and the 2025 Dark World to the Swampland Workshop for their hospitality during the completion of this work. The work of ET is supported in part by the ERC Starting Grant QGuide-101042568 - StG 2021.
\end{acknowledgements}

\appendix
\section{Isometries of Bryant-Salamon Orbifolds}\label{app:isom}
The isometry group for the orbifold $\Sigma(S^3_b/\mathbb{Z}_N)$ is the subgroup of $SU(2)^3\rtimes \mathbb{Z}_2$ generated by elements $g\neq u^k_0$ such that $g^{-1}\mathbb{Z}_Ng=\mathbb{Z}_N$ which is
\begin{equation}\label{eq:cases}
G_{isom.}=\begin{cases}
                         \left(\frac{SU(2)_u}{\mathbb{Z}_2}\times \frac{SU(2)_v\times SU(2)_w}{\mathbb{Z}_2}\right), \;  N=2 & \\
                                      \left(O(2)_u\times\frac{SU(2)_v\times SU(2)_w}{\mathbb{Z}_2}\right), \;  N>2 \; \mathrm{even} & \\
                                      \left(\frac{\widetilde{O(2)}_u\times SU(2)_v\times SU(2)_w}{\mathbb{Z}_2}\right), \;  N \; \mathrm{odd} &
                                    \end{cases}
\end{equation}
Explaining our notation, the $O(2)_u\subset SU(2)_u$ subgroup includes the $U(1)_u\subset SU(2)_u$ subgroup with elements $\mathrm{diag}(e^{i\phi},e^{-i\phi})$ as well as
\begin{equation}\label{eq:ud}
  u_D:=\begin{pmatrix}
         0 & -1 \\
         1 & 0
       \end{pmatrix}
\end{equation}
which acts as an outer automorphism on $U(1)_u$. When $N$ is even $u_D^2=\mathrm{diag}(-1,-1)=u^{N/2}_0$ which implies that $u_D$ is an order-2 isometry, while if $N$ is odd, $u_D^2=\mathrm{diag}(-1,-1)\neq u^k_0$ for any $k$ so is a an order-4 isometry. This is the reason that the group $\widetilde{O(2)}:=(U(1)\rtimes \mathbb{Z}_4)/\mathbb{Z}_2$ appears in the $N$ odd case, where the semidirect product is defined by the non-trivial homomorphism $\mathbb{Z}_4\rightarrow \mathrm{Out}(U(1))=\mathbb{Z}_2$ and the $\mathbb{Z}_2$ quotient identifies $-1\in U(1)$ with $2 \; \mathrm{mod} \; 4\in \mathbb{Z}_4$.

In the above coordinates, one can take the minimal length torsional 1-cycle $\gamma_1$ to be located along the hypersurface intersection
\begin{equation}\label{eq:gamma1hyper}
  \gamma_1=\{r=r_a\}\cap \{z_{2a}=0\},
\end{equation}
and from the metric the length of $\gamma_1$ is
\begin{equation}\label{eq:gammalength}
  L_\gamma=\frac{\pi r_a}{N\sqrt{6}}.
\end{equation}

Isometry groups for the D- and E-type quotients are more straightforward to describe because the subgroup of $SU(2)_u$ which commutes with the non-abelian $\Gamma_{\mathfrak{g}}$ is smaller and is given by the symmetries of the Dynkin diagram of $\mathfrak{g}$ (see for instance \cite{Witten:2001uq}). In the case of $\Gamma_{D_{4+k}}$, which can be presented by
\begin{equation}
   D_{4+k}=\Biggl \langle \begin{pmatrix}
        \zeta_{4+k} & 0 \\
        0 & \zeta^{-1}_{4+k}
    \end{pmatrix}, \; \begin{pmatrix}
        0 & -1 \\
        1 & 0
    \end{pmatrix}\Biggr \rangle,
\end{equation}
then there is a $\mathbb{Z}_2$ outer automorphism given by acting with
\begin{equation}\label{eq:elementrho}
     \begin{pmatrix}
        \zeta_{2(4+k)} & 0 \\
        0 & \zeta^{-1}_{2(4+k)}
    \end{pmatrix}
\end{equation}
by conjugation on $D_{4+k}$ elements \footnote{For $D_4$ there is also an order 3 outer automorphism due to triality but this will not be relevant.}. Meanwhile, $\Gamma_{E_6}$ is simply $D_4$ adjoined with an element
\begin{equation}
    \frac{1}{\sqrt{2}}\begin{pmatrix}
        \zeta^{-1}_8 & \zeta^{-1}_8 \\
        -\zeta_8 & \zeta_8
    \end{pmatrix}.
\end{equation}
There is a $\mathbb{Z}_2$ outer automorphism in this case which is just the result of acting by $\mathrm{diag}(\zeta_8, \zeta^{-1}_8)$ with conjugation. The group $(SU(2)_v\times SU(2)_w)/\mathbb{Z}_2$ is still a part of the isometry group just as with the $\mathbb{Z}_{N}$ orbifolds for even $N>2$.

\bibliography{confinementstring}

\end{document}